# Spontaneous and Directed Symmetry Breaking in the Formation of Chiral Nanocrystals

Uri Hananel, Assaf Ben-Moshe[†], Haim Diamant, Gil Markovich*

*School of Chemistry, Raymond and Beverly Sackler Faculty of Exact Sciences, Tel Aviv University, Tel Aviv 6997801, Israel.*

* Email address: gilmar@post.tau.ac.il

[†]Present address: Department of Chemistry, University of California—Berkeley, Berkeley, California 94720, United States.

**The homochirality of biomolecules remains one of the outstanding puzzles concerning the beginning of life. Chiral amplification of a randomly perturbed racemic mixture of chiral molecules is a well-accepted prerequisite for all routes to biological homochirality[1–6]. Some models have suggested that such amplification occurred due to asymmetric discrimination of chiral biotic/prebiotic molecules when they adsorbed onto crystalline surfaces[7–9]. While chiral amplification has been demonstrated on surfaces of both chiral and achiral crystals[10–15], the mechanism that would produce an enantiomeric imbalance in the chiral surfaces themselves has not been addressed. Here we report strong chiral amplification in the colloidal synthesis of intrinsically chiral lanthanide phosphate nanocrystals, quantitatively measured via the circularly polarized luminescence of the lanthanide ions within the nanocrystals. The amplification involves spontaneous symmetry breaking into either left- or right-handed nanocrystals below a critical temperature. Furthermore, chiral tartaric acid molecules in the solution act as an external "chiral field", sensitively directing the amplified nanocrystal handedness through a discontinuous transition between left- and right-handed excess. These characteristics suggest a conceptual framework for chiral amplification, based on the statistical thermodynamics of**

**critical phenomena, which we use to quantitatively account for the observations. Our results demonstrate how chiral minerals with high enantiomeric excess could have grown locally in a primordial racemic aqueous environment.**

The influence of biomolecules on crystal growth has been studied for decades[16]. Yet, particularly, biomolecules affecting intrinsically chiral inorganic crystals[17] have received little attention. Such crystals could be relevant to the question of biological homochirality, as their surface is naturally chiral. Only recently were these interactions studied in the context of control over the handedness of intrinsically chiral nanocrystals (NCs)[18–20]. Those NCs, however, did not display chiral amplification, and are not biologically relevant. We therefore turned our attention to intrinsically chiral lanthanide phosphate hydrate crystals, which are biocompatible[21] and were expected to display strong circularly polarized luminescence (CPL), as chiral lanthanide complexes do[22,23]. Chiral $Eu^{3+}$-doped $TbPO_4·H_2O$ rod-shaped NCs (see Fig. 1) were prepared from an aqueous solution of lanthanide and phosphate ions in the presence of tartaric acid (TA), at a pH value of ~2 and a range of temperatures within 40-100°C. TA is a chiral biomolecule, which strongly coordinates the lanthanide cations via its carboxylate groups[24], but is not present inside the NCs' lattice.

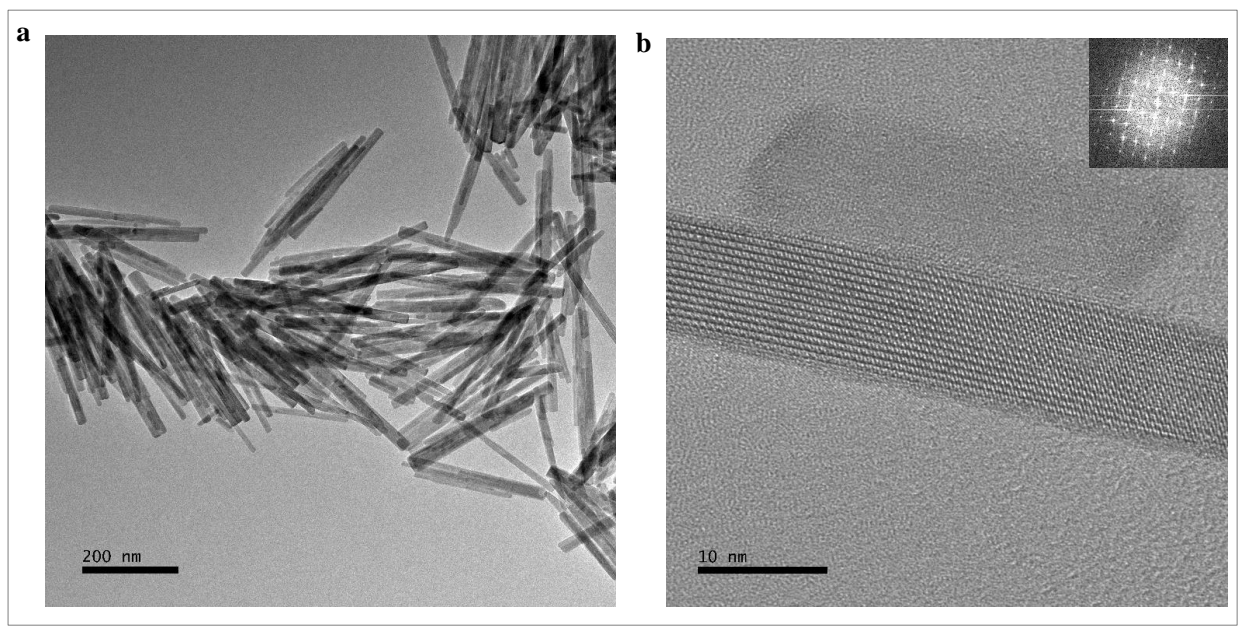

**Fig. 1: Electron microscopy imaging.** TEM images of the synthesized NCs at two magnifications. Inset in **b** is the FFT of the main panel.

Powder x-ray diffraction (Supplementary Fig. 1) verified the chiral space group of the NCs ($P3_121$ or its enantiomorph $P3_221$), in agreement with the literature on similar NCs[21,25]. High resolution transmission electron microscopy (TEM) images indicated single crystals without apparent extended defects (Fig. 1b). As domains of the two enantiomorphic space groups within the same NC must be separated by visible twin defect planes, we infer that each NC is a single crystal of a certain handedness. Further evidence for the single handedness of individual NCs was found in a seeding experiment described in Supplementary Fig. 2.

The luminescence spectrum of the NCs (Fig. 2a) exhibits the sharp bands characteristic of lanthanide emission[26]. Emission from the surface, or from precursor contaminants, is either quenched or non-existing. For a detailed discussion see Supplementary Fig. 3.

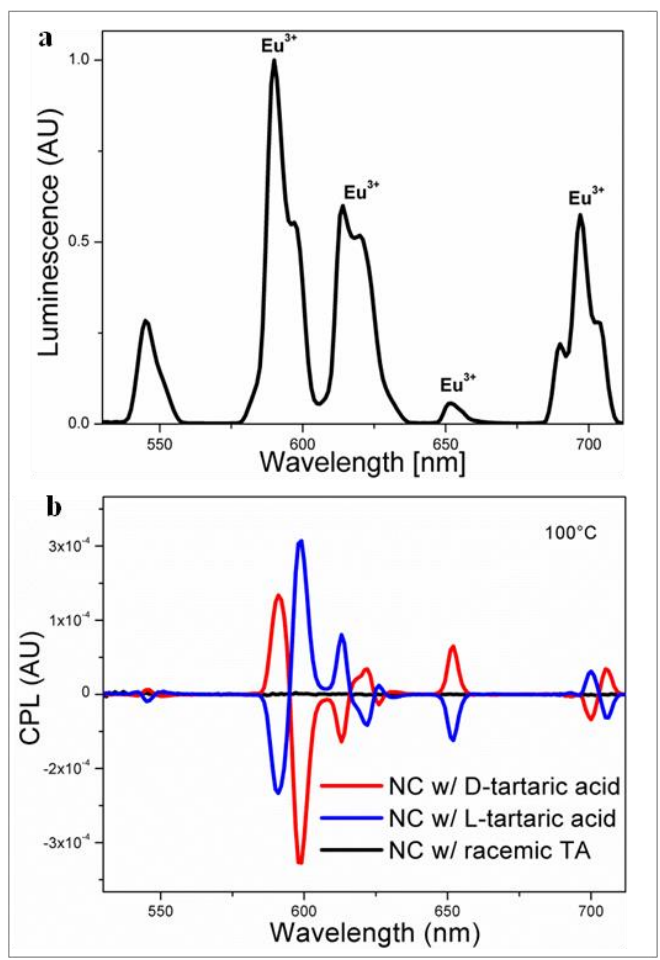

**Fig. 2: The emission of the NCs. a.** Luminescence spectrum (excitation wavelength $\lambda_{ex} = 365$ nm) of $Eu^{3+}$-doped $TbPO_4 \cdot H_2O$ NCs synthesized at 100°C. **b.** CPL spectra ($\lambda_{ex} = 365$ nm) of the NCs synthesized at 100°C with enantiopure D-TA, L-TA, and a racemic mixture of the TA. The emission at and above 590 nm stems solely from $Eu^{3+}$ ions inside the $TbPO_4 \cdot H_2O$ NCs.

CPL is a measure of the degree of circular polarization in luminescence from chiral systems[22,23], and is defined as: CPL $\equiv I_L - I_R$, where $I_L$ is the emitted left-handed circularly polarized light and $I_R$ is the right-handed one. TbPO$_4$·H$_2$O NCs synthesized in the presence of pure right-handed (D-) TA display several strong CPL lines (Fig. 2b), indicating that the lanthanide ions are in a chiral environment and that there is a substantial enantiomeric excess (*ee*) of NCs. We label the NC enantiomers in excess (made with D-TA) as Δ-NCs. As expected, changing to left-handed (L-) TA gives rise to excess NCs with the opposite handedness (labeled as Λ-NCs), resulting in sign inversion of the CPL spectrum. Likewise, using a racemic mixture of TA for the synthesis at 100° results in zero CPL (Fig. 2b).

In reporting CPL data it is convenient to use the dissymmetry factor, $g_{lum} = 2\frac{I_L - I_R}{I_L + I_R}$. This normalization eliminates the dependence on concentration and instrumentation, and is proportional to the *ee* of the emitting species. Therefore, if $g_{lum}$ of an enantiopure sample is known, it can be used as a "ruler" to quantify the *ee* of different preparations of the same material. The largest $g_{lum}$ appears at the weakest (electric dipole-forbidden[22,23]) transition at 651 nm, and its value is ~0.4, which is comparable to the highest values found for enantiopure chiral lanthanide complexes[22,23,27]. For other values of the NC $g_{lum}$ at enantiopure conditions, see Supplementary Table 1.

To estimate the magnitude of the NC *ee*, the NCs were synthesized with different ratios of the two TA enantiomers. The TA *ee* is defined as $h = (x_L - x_D)/(x_L + x_D)$, where $x_D$ and $x_L$ are the molar fractions of D-TA and L-TA, respectively. Fig. 3 displays the dependence of the absolute value of the NCs' $g_{lum}$ on the absolute value of the TA *ee* used for their synthesis. Remarkably, the NCs' |$g_{lum}$| vs. |*h*| curve is non-linear, saturating at the relatively low *h* value of ~0.4.

Unlike an earlier study, where the absorption *g*-factor (which is proportional to the NC *ee*) was found to change linearly with $h$[19], here, the saturation of $g_{lum}$ at intermediate *h* values indicates that the maximum excess *(NC ee* ≈ 1) has been reached. This enantiopurity was confirmed by single particle CPL microscopy measurements[28]. To our best knowledge, this is the first demonstration of enantiopurity in a chiral NC synthesis. Thus, the specific limiting $g_{lum}$ value for each emission line defines a ruler for converting $g_{lum}$ into NC *ee*. The saturation of $g_{lum}$ seems special to syntheses with TA. When a similar chiral dicarboxylic acid (glutamic acid) was used instead of TA at the exact same synthesis conditions, the NC *ee* changed sub-linearly with *h* and did not saturate (Fig. 3 inset). The much weaker effect of glutamic acid demonstrates the significance of molecular structural details in controlling the handedness of the NCs. The TA-based synthesis clearly exhibits *chiral amplification* – a small *ee* in the reactants (TA ligands) is amplified to a much greater *ee* in the resulting products (NCs). Note that a racemic TA mixture leads to the formation of a racemic NC mixture when synthesized at 100°C. As we present shortly, a strikingly different result is obtained at lower temperatures.

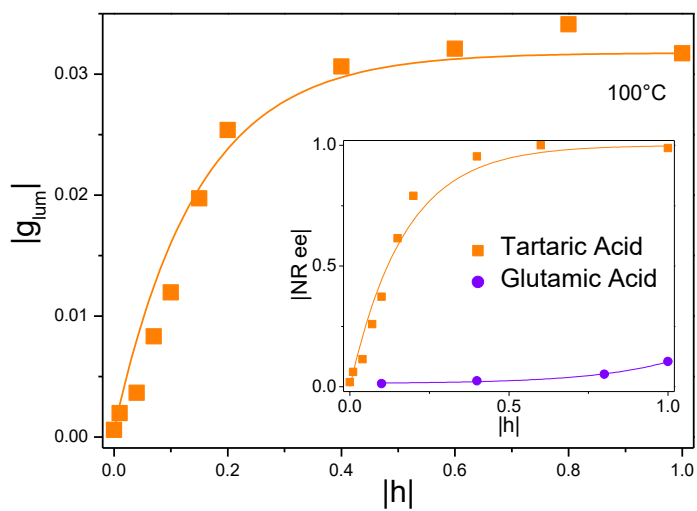

**Fig. 3: NC dissymmetry as a function of the TA enantiomeric excess.** The dissymmetry factor at 699 nm for the NCs synthesized with different *h* values. Both $g_{lum}$ and *h* are shown in absolute value for simplicity. The inset compares the behavior of the TA-based synthesis to that of a glutamic acid-based one, where both are normalized by the saturation value of $g_{lum}$ observed for the TA-based synthesis. All syntheses in this figure were performed at 100°C. The solid lines are added as guides to the eye.

Once a ruler for estimating the NC *ee* has been set (normalizing all $g_{lum}(h)$ by the $g_{lum}$ value at saturation), we can turn to study the NC *ee* obtained from syntheses under different conditions. On lowering the temperature to 50 and 40°C, we observed an increase in the slope of the NC *ee* vs. *h* curve (Fig. 4a-c), indicating increased chiral amplification.

Focusing on the racemic case (*h* = 0), we reach one of our key observations: At the lower temperatures, for racemic TA conditions, a NC *ee* of large magnitude and arbitrary handedness appears (Fig. 4b,c for *h* = 0). At 40°C this spontaneous NC *ee* reached values as large as 0.8. Moreover, the same was observed in syntheses of the NCs in the complete absence of TA (Fig.

4d), where similar rod-shaped NCs were formed. In both cases, the average NC *ee* over many syntheses approached zero, as is expected by symmetry. These two sets of experiments (racemic TA and no TA, in either case vanishing *h*) indicate the occurrence of spontaneous symmetry breaking at sufficiently low temperatures. This implies "infinite" chiral amplification – large *ee* in the products associated with zero *ee* in the reactants. The symmetry breaking probably originates in a small random NC *ee* formed at an early synthesis stage. Similar experiments performed without stirring led to no symmetry breaking (Supplementary Fig. 4). This suggests that the underlying mechanism might be *secondary nucleation*, where an existing crystal surface facilitates the formation of new crystals of the same handedness, thus introducing a positive correlation between same-handedness NCs and leading to chiral amplification. Previously, such a mechanism presumably allowed growth (and ripening) of chiral $NaClO_3$ crystals of near enantiopurity in the absence of any chiral moiety, as demonstrated by Kondepudi *et al.*[30] and more recently by Viedma and coworkers[2,31]. In this scenario, stirring should promote the release of secondary seeds from their progenitor NCs[32].

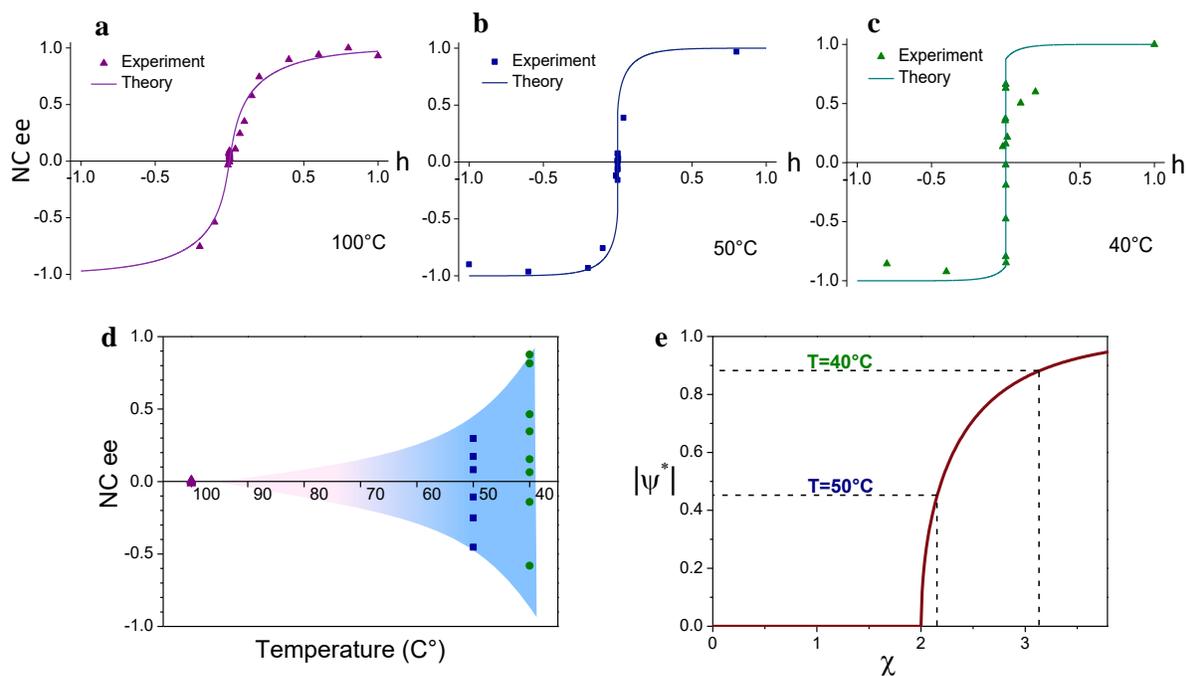

**Fig. 4**: **NC *ee* as a function of the chiral field above and below the critical temperature. a-c.** NC *ee* as a function of *h* for three different reaction temperatures. The points are the experimental data and the lines are the fitted model curves. Here the positive and negative values correspond to opposite handedness, with a positive *h* value yielding, for the 100°C synthesis, a positive NC *ee* value. **d.** Spontaneous NC *ee* values measured in the absence of TA for the three reaction temperatures. **e.** The theoretical

spontaneously formed NC *ee* ($\psi^*$) as a function of the correlation parameter $\chi$, along with the fitted $\chi$ values at two experimental temperatures. The free fitting parameters used in panels **a-c** are: $\varepsilon^+$ = -0.2 eV [29]; $\varepsilon^-$ = 0.99$\varepsilon^+$; M = 2000.

The findings described above bear the signatures of a critical phenomenon[33], with the NC *ee* being the order parameter, and *h* (the TA *ee*) being a symmetry-breaking external field. Above a critical temperature $T_c$ (somewhere in the range 50–100°C), the NC *ee* as a function of *h* is a continuous curve, and at *h* = 0 the NC *ee* vanishes (Fig. 4a). Below $T_c$, as *h* changes from negative to positive values, the NC *ee* undergoes a discontinuous (first-order) transition between left- and right-handedness (Fig. 4b,c). At *h* = 0 the symmetry is spontaneously broken into a finite, either positive or negative NC *ee* (At $T = T_c$ there should be a second-order transition, which was not captured in the experiment). Another signature of a first-order transition is the possibility to have a metastable state beyond the transition point (e.g. hysteresis in ferromagnets; supercooling/superheating in solid-liquid phase transitions). Our system exhibits this property as well – at the lower temperatures we observed also NC *ee* of the "wrong" handedness, i.e., opposite to that of *h*; see Supplementary Fig. 5.

Many models of chiral amplification, starting from the pioneering work by Frank[34], have relied on catalytic chemical kinetics[1]. Based on the aforementioned observations, we propose an alternative approach, relying on the statistical mechanics of critical phenomena. This description fits our results well, and may be found relevant to other chiral systems[35]. We assume the following: (a) NC growth is sufficiently slow to allow the adsorption of TA molecules on NC surfaces to be in quasi-equilibrium at all times. The time scale of NC growth ranges from seconds at 100°C to hours at 40°C, which at any rate is much longer than the adsorption relaxation time. (b) The NCs are positively correlated, i.e., the presence of NCs of a certain handedness facilitates the formation of more of the same handedness, even in the absence of the directing TA. Such a correlation may be brought about by secondary nucleation. (c) The adsorption of TA molecules of a certain handedness onto a NC surface of a fitting handedness is slightly more favorable than their adsorption on a surface of the opposite NC handedness. Such a difference in adsorption free energies is to be expected, and has been previously reported[13,36]. (d) The total number of NCs is fixed. This constraint is set by the limited amount of precursor ions in the synthesis vessel and the relatively uniform sizes of the observed rod-shaped NCs. Based on these experimentally supported assumptions, we construct the following phenomenological model.

We begin by considering a total number $N$ of NCs, in the absence of TA, divided into $N_\Delta = yN$ right-handed NCs and $N_\Lambda = (1-y)N$ left-handed ones. The NC *ee* is $\psi = (N_\Delta - N_\Lambda)/N = 2y - 1$. If the NCs nucleated randomly and independently, the probability of having a given partition into $N_\Delta$ and $N_\Lambda$ would be $2^{-N}N!/(N_\Delta! N_\Lambda!)$, which is peaked at $\psi = 0$. However, the positive correlation between same-handedness nuclei, represented here by a single parameter $\chi$, penalizes opposite handedness by a factor $e^{-N\chi y(1-y)}$ [37]. The resulting probability distribution, $P_0(\psi)$, is peaked at a NC *ee* that satisfies the equation, $\psi = \tanh(\chi\psi/2)$. This equation describes a critical behavior with spontaneous symmetry breaking. For $\chi < 2$ the only solution is $\psi = 0$, i.e., a single peak at zero excess. For $\chi > 2$ two nonzero solutions appear, $\pm\psi^*$, corresponding to a distribution with two peaks at nonzero left- and right-handed spontaneous excesses. The magnitude of the spontaneous excess $|\psi^*|$ as a function of $\chi$ is shown in Fig. 4e. Comparing it with the magnitude of spontaneous NC *ee* measured for 40° and 50°C in the absence of TA, we get the estimates $\chi(40°) \approx 3.1$ and $\chi(50°) \approx 2.15$. Since 100°C lies above the critical temperature, $\chi(100°) < 2$. The decrease of correlation with temperature is the generally expected trend. If the correlation mechanism is secondary nucleation, it might be related to an increase in primary nucleation with temperature, which would make secondary nucleation relatively less significant.

Next, we consider the effect of TA. The surface of each NC is assumed to contain $M$ independent adsorption sites, each of which can be either vacant, occupied by a TA molecule of a particular handedness fitting the NC handedness (adsorption energy $\varepsilon^+$), or a TA molecule of opposite handedness (adsorption energy $\varepsilon^- > \varepsilon^+$). We define the four factors, $\zeta_i^j \equiv e^{(\mu_i - \varepsilon^j)/k_B T}$ ($i = L, D$, $j = +, -$), where $\mu_i$ is the chemical potential of the *i*-handed TA molecules in the solution, $k_B$ the Boltzmann constant, and $T$ the temperature. Assuming an ideal solution, we have $\zeta_i^j = x_i e^{-\varepsilon^j/k_B T}$, where $x_i$ is the molar fraction of *i*-handed TA. We define the TA *ee* as the external "chiral field", $h \equiv (\zeta_L^+ - \zeta_D^+)/(\zeta_L^+ + \zeta_D^+) = (\zeta_L^- - \zeta_D^-)/(\zeta_L^- + \zeta_D^-) = (x_L - x_D)/(x_L + x_D)$. The additional degrees of freedom introduced by the adsorption affect the probability distribution of $\psi$, the NC *ee*, as follows. The partition function of a single adsorption site on a Δ-NC is $q_\Delta = 1 + \zeta_D^+ + \zeta_L^-$, and that of a site on a Λ-NC is $q_\Lambda = 1 + \zeta_L^+ + \zeta_D^-$. These give rise to the statistical weights $(q_i)^{MN_i}$, which multiply the TA-free probability distribution. Using the relations between $N_{\Delta,\Lambda}$ and $\psi$, and between $\zeta_i^j$ and $h$, we have, finally,

$$P(\psi, h) \propto P_0(\psi) \times$$

$$\left[1 + \frac{1}{2}x(1+h)e^{-\frac{\varepsilon^+}{k_BT}} + \frac{1}{2}x(1-h)e^{-\frac{\varepsilon^-}{k_BT}}\right]^{MN(1+\psi)/2} \left[1 + \frac{1}{2}x(1-h)e^{-\frac{\varepsilon^+}{k_BT}} + \frac{1}{2}x(1+h)e^{-\frac{\varepsilon^-}{k_BT}}\right]^{MN(1-\psi)/2}$$

where $x \equiv x_L + x_D$ is the total molar fraction of TA. For $x = 0$, $P(\psi, h)$ reduces to the TA-free distribution $P_0(\psi)$. The same is true for $h = 0$ (racemic TA mixture) and for $\varepsilon^+ = \varepsilon^-$ (no preference for same-handedness adsorption), whereby the factor due to adsorption becomes a constant independent of $\psi$, as demanded by symmetry.

For $\chi < 2$ the distribution is single-peaked, as in the TA-free case, and changing $h$ above or below 0 shifts the location of the peak to positive or negative values of $\psi$, thus directing the NC *ee*; see the fit in Fig. 4a. For $\chi > 2$ and $h = 0$, the distribution has two peaks of equal heights at opposite values of $\psi$, as in the TA-free case. As soon as $h$ slightly deviates up or down from 0, the corresponding positive or negative peak becomes superior and determines the excess, an extreme amplification of the NC chirality, directed by the TA; see the fits in Fig. 4b,c. For small values of $|h|$ the inferior peak still exists, allowing for the occasional appearance of "supercooling", i.e., *ee* of opposite signs for the TA and NC (Supplementary Fig. 5). At higher $h$ values the inferior peak disappears, and an equal sign of $\psi$ and $h$ is the only possibility. The theoretical fits in Fig. 4a-c use three free parameters: $\varepsilon^-$, $\varepsilon^+$, and M. We note that fits of similar quality could be obtained using other parameter values. Thus, at present, such fits cannot be used to extract the system's parameters reliably. They merely demonstrate full consistency between experiment and theory.

The extreme sensitivity of the NC *ee* to the presence of TA at the lower temperatures is a consequence of the criticality in NC formation. The results are strongly dependent also on the difference in adsorption energies, $\varepsilon^+ - \varepsilon^-$. The reason for this sensitivity is the large number of adsorption sites $M$ on each NC; what actually matters is the adsorption energy difference for an entire NC, $M(\varepsilon^+ - \varepsilon^-)$. Thus, since $M$ is at least of order $10^2$–$10^3$, a minute contrast in the molecular adsorption energies is sufficient to produce strong TA-directed chiral amplification even above the critical temperature, where NC correlations are less dominant. The size-intensified role of adsorption might explain also the sensitivity to the molecular structure of the adsorbate (cf. Fig. 3 inset).

We have presented for the first time directed, as well as spontaneous, non-linear chiral amplification in the formation of inorganic chiral crystals. As this amplification is guided by a biomolecule, and occurs in a mineral which is relatively abundant, catalytically active[38–40], and biocompatible[21], it is all the more relevant to biological homochirality. The behavior of the system displays the features of a phase transition. Intrinsically chiral inorganic NCs could therefore serve as excellent model systems to test the characteristics of chiral symmetry breaking as a critical phenomenon. We also show here that the competition between chiral influences and random effects, thought to have been relevant for the origins of homochirality, is relevant also for the growth of inorganic, biologically relevant, chiral minerals.

This work also highlights the huge, and as of yet untapped, potential of enantioselective synthesis of intrinsically chiral inorganic crystals. From homochirality to chiral surfaces and biological probes, there is much to be learned from such systems.

**METHODS**

**Synthesis of NCs.** Terbium chloride hexahydrate (99.9%) and europium chloride hexahydrate (99.9%) were purchased from Fisher Scientific. Sodium dibasic phosphate (≥99.0%), D-tartaric acid (99%), L-tartaric acid (≥99.5%), L-glutamic acid (≥99%), D-glutamic acid (≥99%) and (+)-camphoric acid (99%) were purchased from Sigma-Aldrich. Hydrogen chloride (32%) was purchased from BioLab. All chemicals were used without further purification. All water used was ultrapure (18.2 MΩ), obtained from a Millipore Direct-Q3 UV.

All precursor chemicals were diluted in water to 100 mM stock solutions, except for the hydrogen chloride solution which was diluted in water to 1M.

*Solution **A** preparation*. In a round-bottom flask, to 20 mL water we added, at the order as listed: 570 μL of $TbCl_3·6H_2O$ stock solution, 30 μL of $EuCl_3·6H_2O$ stock solution, 300 μL of HCl 1M and 600 μL of TA solution, with the ratio between D- and L-tartaric acid set as desired. The solution was heated to the desired synthesis temperature (40-100°C) while stirring.

*Solution **B** preparation*. in a glass vial, to 2 mL of water we added 1.2 mL of $Na_2HPO_4$ solution and 300 μL of HCl 1M and the solution was heated to the synthesis temperature.

The solutions were allowed 30 minutes to equilibrate at the desired temperature, following which solution **B** was rapidly added to solution **A** while stirring. The addition of the phosphate solution to the $Ln^{3+}$ solution prompted the precipitation of $Eu^{3+}$ doped $TbPO_4 \cdot H_2O$ NCs, which caused the solution to become turbid over time, the rate depending on the temperature of the synthesis and the TA enantiomeric excess. NC samples were purified via centrifugation at 3300 RCF for 10 minutes and re-dissolution of the precipitate in water.

**Luminescence and circularly polarized luminescence.** All measurements were performed in a standard home-built CPL system using a photoelastic modulator and lock-in detection, built according to the general setup described in ref. 23. Samples were excited using a high-power near-UV LED light source (Hamamatsu, 365 nm, 1.4 $W/cm^2$). Light emitted from the samples was collected at 90° to the excitation, with both excitation and emission edgepass filters to minimize light leakage from the excitation source, passed through the monochromator and detected by a photomultiplier. The NC solutions were magnetically stirred in the cuvette during the CPL measurements to ensure solution uniformity.

**Structural characterization.** TEM imaging was performed using a FEI Tecnai F20 FEG-TEM. TEM samples were prepared on carbon-coated 300-mesh Cu grids. Samples for powder X-ray diffraction (XRD) were prepared by centrifuging at 3300 RCF for 10 minutes, removing the supernatant and drying overnight at room temperature under vacuum. XRD data was collected in a symmetric Bragg-Brentano geometry with CuKα radiation on Bruker D8 Discover Θ:Θ X-ray diffractometer equipped with a one-dimensional LynxEye XE detector based on compound silicon strip technology.

**REFERENCES**

1. Blackmond, D. G. The origin of biological homochirality. *Cold Spring Harb. Perspect. Biol.* **2,** a002147 (2010).
2. Viedma, C. Chiral Symmetry Breaking During Crystallization: Complete Chiral Purity Induced by Nonlinear Autocatalysis and Recycling. *Phys. Rev. Lett.* **94,** 065504 (2005).
3. Klussmann, M. *et al.* Thermodynamic control of asymmetric amplification in amino acid catalysis. *Nature* **441,** 621–623 (2006).
4. Perry, R. H., Wu, C., Nefliu, M. & Cooks, R. G. Serine sublimes with spontaneous chiral amplification. *Chem. Commun.* 1071–1073 (2007).
5. Soai, K., Shibata, T., Morioka, H. & Choji, K. Asymmetric autocatalysis and

amplification of enantiomeric excess of a chiral molecule. *Nature* **378,** 767–768 (1995).
6. Weissbuch, I., Bolbach, G., Leiserowitz, L. & Lahav, M. Chiral amplification of oligopeptides via polymerization in two-dimensional crystallites on water. *Orig. Life Evol. Biosph.* **34,** 79–92 (2004).
7. Hazen, R. M. & Sholl, D. S. Chiral selection on inorganic crystalline surfaces. *Nat. Mater.* **2,** 367–374 (2003).
8. Hazen, R. M. & Sverjensky, D. A. Mineral Surfaces, Geochemical Complexities, and the Origins of Life. *Cold Spring Harb. Perspect. Biol.* **2,** a002162 (2010).
9. Ma, W. *et al.* Chiral Inorganic Nanostructures. *Chem. Rev.* **117,** 8041–8093 (2017).
10. Hazen, R. M., Filley, T. R. & Goodfriend, G. A. Selective adsorption of L- and D-amino acids on calcite: Implications for biochemical homochirality. *Proc. Natl. Acad. Sci.* **98,** 5487–5490 (2001).
11. Soai, K. *et al.* D- and L-Quartz-Promoted Highly Enantioselective Synthesis of a Chiral Organic Compound. *J. Am. Chem. Soc.* **121,** 11235–11236 (1999).
12. Yun, Y. J. & Gellman, A. J. Adsorption-induced auto-amplification of enantiomeric excess on an achiral surface. *Nat. Chem.* **7,** 520–525 (2015).
13. Gellman, A. J. *et al.* Superenantioselective chiral surface explosions. *J. Am. Chem. Soc.* **135,** 19208–19214 (2013).
14. Fasel, R., Parschau, M. & Ernst, K.-H. Amplification of chirality in two-dimensional molecular lattices. *Nature* **439,** 449–452 (2006).
15. Ortega Lorenzo, M., Baddeley, C. J., Muryn, C. & Raval, R. Extended surface chirality from supramolecular assemblies of adsorbed chiral molecules. *Nature* **404,** 376–379 (2000).
16. Weiner, S. & Addadi, L. Crystallization Pathways in Biomineralization *Annu. Rev. Mater. Res.* **41,** 21–40 (2011).
17. Dryzun, C. & Avnir, D. On the abundance of chiral crystals. *Chem. Commun.* **48,** 5874–5876 (2012).
18. Ben-Moshe, A. *et al.* Enantioselective control of lattice and shape chirality in inorganic nanostructures using chiral biomolecules. *Nat. Commun.* **5,** 4302 (2014).
19. Ben-Moshe, A., Govorov, A. O. & Markovich, G. Enantioselective synthesis of intrinsically chiral mercury sulfide nanocrystals. *Angew. Chemie - Int. Ed.* **52,** 1275–1279 (2013).
20. Wang, P.-P., Yu, S.-J., Govorov, A. O. & Ouyang, M. Cooperative expression of atomic chirality in inorganic nanostructures. *Nat. Commun.* **8,** (2017).
21. Di, W. *et al.* Photoluminescence, cytotoxicity and in vitro imaging of hexagonal terbium phosphate nanoparticles doped with europium. *Nanoscale* **3,** 1263–9 (2011).
22. Carr, R., Evans, N. H. & Parker, D. Lanthanide complexes as chiral probes exploiting circularly polarized luminescence. *Chem. Soc. Rev.* **41,** 7673–86 (2012).
23. Zinna, F. & Di Bari, L. Lanthanide circularly polarized luminescence: Bases and applications. *Chirality* **27,** 1–13 (2015).
24. Bünzli, J.-C. G., Piguet, C., Bunzli, G., J.-C. & & Piguet C. Taking advantage of luminescent lanthanide ions. *Chem. Soc. Rev.* **34,** 1048 (2005).
25. Fang, Y.-P. *et al.* Systematic Synthesis and Characterization of Single-Crystal Lanthanide Orthophosphate Nanowires. *J. Am. Chem. Soc.* **125,** 16025–16034 (2003).
26. Bunzli, J. G. & Eliseeva, S. V. *Basics of Lanthanide Photophysics. Springer Series on Fluorescence* (2011).

27. Petoud, S. *et al.* Brilliant Sm, Eu, Tb, and Dy chiral lanthanide complexes with strong circularly polarized luminescence. *J. Am. Chem. Soc.* **129,** 77–83 (2007).
28. Vinegrad, E., Hananel, U., Markovich, G. & Cheshnovsky, O. Determination of handedness in single chiral nanocrystals via circularly polarized luminescence. *arXiv:1808.05626* (2018).
29. The energy of TA adsorption to the NCs was estimated to be ~20 kJ/mol from isothermal titration calorimetry experiments performed by Liora Werber and Itzhak Mastai, Bar Ilan University, private communication.
30. Kondepudi, D. K., Kaufman, R. J. & Singh, N. Chiral Symmetry Breaking in Sodium Chlorate Crystallizaton. *Science* **250,** 975–976 (1990).
31. Sögütoglu, L.-C., Steendam, R. R. E., Meekes, H., Vlieg, E. & Rutjes, F. P. J. T. Viedma ripening: a reliable crystallisation method to reach single chirality. *Chem. Soc. Rev.* 6723–6732 (2015).
32. Kondepudi, D. K., Bullock, K. L., Digits, J. A. & Yarborough, P. D. Stirring Rate as a Critical Parameter in Chiral Symmetry Breaking Crystallization. *J. Am. Chem. Soc.* **117,** 401–404 (1995).
33. Binney, J. J., Dowrick, N. J., Fisher, A. J. & Newman, M. E. J. *The Theory of Critical Phenomena*. (Oxford University Press, 1992).
34. Frank, F. C. On spontaneous asymmetric synthesis. *Biochim. Biophys. Acta* **11,** 459–463 (1953).
35. Relating homochirality to a condensation transition was suggested a few decades ago within a rather exotic scenario based on the electroweak interaction: Salam, A. The role of chirality in the origin of life. *J. Mol. Evol.* **33,** 105–113 (1991).
36. Baldanza, S., Ardini, J., Giglia, A. & Held, G. Stereochemistry and thermal stability of tartaric acid on the intrinsically chiral Cu{531} surface. *Surf. Sci.* **643,** 108–116 (2016).
37. The resulting probability distribution is equivalent to the Flory-Huggins partition function of polymer solutions, whereby χ is analogous to Flory's χ parameter: Rubinstein, M. & Colby, R. H. *Polymer Physics*. (Oxford University Press, 2003).
38. Onoda, H., Nariai, H., Moriwaki, A., Maki, H. & Motooka, I. Formation and catalytic characterization of various rare earth phosphates. *J. Mater. Chem.* **12,** 1754–1760 (2002).
39. Devi, G. S., Giridhar, D. & Reddy, B. M. Vapour phase O -alkylation of phenol over alkali promoted rare earth metal phosphates. *J. Mol. Catal. A Chem.* **181,** 173–178 (2002).
40. Nguyen, T. T. N., Bellière-Baca, V., Rey, P. & Millet, J. M. M. Efficient catalysts for simultaneous dehydration of light alcohols in gas phase. *Catal. Sci. Technol.* **5,** 3576–3584 (2015).

**ACKNOWLEDGEMENTS**

This work was supported by Israel Science Foundation grant no. 507/14. We thank Eitam Vinegrad for his help with the design of the CPL instrumentation. Isothermal titration calorimetry measurements of TA adsorption on the NCs by Liora Werber and Yitzhak Mastai are gratefully acknowledged.

**AUTHOR INFORMATION**
Author contributions:
A.B.M., U.H. and G.M. conceived the experiments. U.H. did all the experimental work. G.M. supervised the project. H.D. conceived the model and did the calculations. All authors contributed to scientific discussions and writing of the manuscript.

Competing interests
The authors declare no competing interests.

Corresponding authors
Correspondence to Gil Markovich.

**SUPPLEMENTARY INFORMATION**

X-ray diffraction data of the NCs, NC seeded growth experiments, calculation of number of TA molecules adsorbed on the seed particles, discussion on the origin of the NCs' CPL, the NCs' $g_{lum}$ values, results of syntheses without stirring, and hysteresis behavior in the NC *ee* vs. *h* curves.

List of abbreviations:

TA = tartaric acid, UC = unit cell, ee = enantiomeric excess

1. <u>Powder XRD</u>

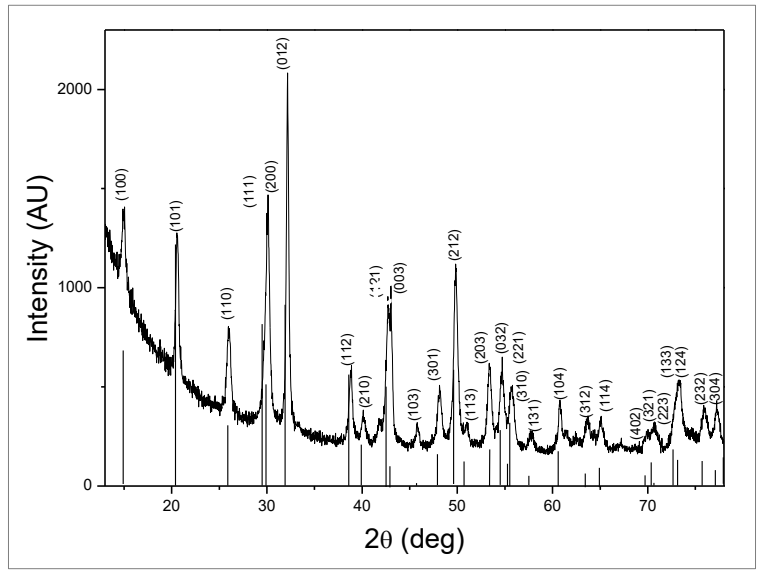

**Figure S1.** Powder XRD pattern of the NCs prepared with TA having ee=0.2, with bulk reference presented as vertical lines, in agreement with the literature (JCPDS No. 20-1244). NCs synthesized with different TA ee displayed the same pattern.

2. NC seeding

The seed NCs are small spherical $Eu^{3+}$-doped $TbPO_4·H_2O$ NCs, produced by the same process as the NC synthesis in the main text, but at room temperature, where the NC formation occurs over the scale of days. The seed NCs were prepared in enantiopure L-TA (enantiopure seeds) and were isolated in early NC growth stages (1 hour after mixing of precursors) by purification from the synthesis mixture. The purification was done via two centrifugation steps. Decreasing amounts of seed particle solution were then added to NC syntheses at 50°C with racemic TA and the CPL of the resultant, fully grown NCs was measured. We note that at this synthesis temperature, the addition of seed particles causes the NCs growth time to be significantly shorter than NC formation without seeds, indicating that the NCs are primarily growing out of the added seed particles. The luminescence and CPL spectra of the seeded NCs matched those of unseeded ones, yet seeded NCs had a broader size distribution (Figure S2). The NC $ee$ could be extracted by normalizing to the $g_{lum}$ value of the NCs grown with $h=1$ at 100°C (Figure 3 in the paper).

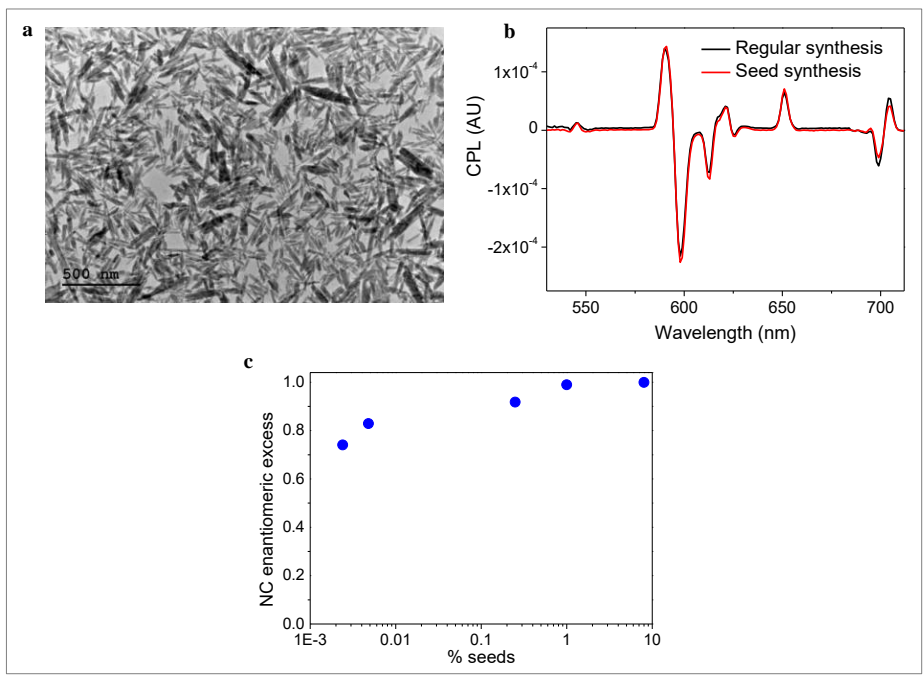

**Figure S2. a.** TEM image of the NCs synthesized with seed particles. **b.** The CPL spectrum of NCs synthesized in regular (no seeds) synthesis in the presence of L-TA, NCs synthesized with racemic TA and 7% NC seeds, the seeds synthesized with L-TA. **c.** The $ee$ of several NC syntheses with different seed quantities (as % from the whole number of particles in such a synthesis).

The amount of added seed particles is reported as % out of the total synthesis volume. As the yield of the seeds is most likely below 100%, the % seeds represents an upper boundary estimate on the amount of seeds, relative to the amount of fully grown NCs. In Figure S2c it can be seen that on adding decreasing amounts of seed particles prepared with L-TA into a precursor mixture containing a racemic TA mixture, the resulting NCs are still close to *ee*=1 (close to enantiopure). <u>This indicates that the NC's handedness is determined at the seeding stage, hence, no change of NC handedness occurs during their growth, which strengthens the claim of single crystal NCs in the unseeded syntheses</u>. This also indicates that the NCs grow out of the preformed seeds with little spontaneous nucleation when the synthesis is done at 50°C.

While it is true that some L-TA is added to the synthesis with the seed particles, thus slightly changing the ee of the TA (*h*) from a perfect racemate, the weak bonding of TA to the NC surface causes much of the adsorbed TA to be released to the solution after centrifugation and dissolution of the precipitate in distilled water. Therefore, two subsequent purifications are expected to remove most of the adsorbed TA. However, even if none of the adsorbed TA is released in these purification steps, the amount of TA added with the seeds is small and cannot account for the large resulting final fully-grown NC ee. For a simple quantitative estimate see the following page.

3. Estimate of the amount of TA adsorbed to seed particles

Number of tartaric acid molecules ($N_{TA}$) added with a certain number of seed particles ($N_{seed}$):

$$(1)\quad N_{TA} = N_{seed} * N_{TA\ adsorbed\ per\ seed}$$

Number of seed particles added, assuming maximal (100% of $Tb^{3+}$ ions reacted to form seeds) yield in the seed synthesis:

$$(2)\quad N_{seed} = N_{Total\ seeds} * \%_{seeds} = \frac{N_{Tb^{3+}\ in\ solution}}{N_{Tb^{3+}\ per\ seed}} * \%_{seeds}$$

Calculation of the number of $Tb^{3+}$ ions per seed particle ($N_{Tb3+\ per\ seed}$) through the number of unit cells per seed particle volume ($N_{UCs\ per\ seed}$):

$$(3)\quad N_{Tb^{3+}\ per\ seed} = N_{UCs\ per\ seed} * N_{Tb^{3+}\ per\ UC} = \frac{V_{seed}}{V_{UC}} * \#_{Tb^{3+}\ per\ UC}$$

Amount of TA adsorbed per seed NC using seed surface area ($S_{seed}$):

$$(4)\quad N_{TA\ adsorbed\ per\ seed} = S_{seed} * N_{TA\ per\ nm^2}$$

Therefore, according to (1)-(4):

$$(5)\quad N_{TA} = \frac{N_{Tb^{3+}\ in\ solution}}{\frac{V_{seed}}{V_{UC}} * N_{Tb^{3+}\ per\ UC}} * \%_{seeds} * S_{seed} * N_{TA\ per\ nm^2}$$

From powder XRD & TEM:

$$(6)\quad S_{seed} \cong 4\pi r^2; V_{seed} \cong \frac{4}{3}\pi r^3; N_{Tb^{3+}\ per\ UC} = 3; V_{UC} = 285.38\ \text{Å}^3; r_{seed}^{average} = 2.5\ nm$$

We assume full coverage, and that the TA density on the surface is:

$$(7)\quad \#_{TA\ per\ nm^2} \approx 4$$

Therefore, the maximal amount of adsorbed TA on a single seed is:

$$(8)\quad N_{TA\ per\ seed} = S_{seed} * N_{TA\ per\ nm^2} \approx 300$$

Furthermore, the synthesis conditions are:

$$(9)\quad \%_{seeds} = 1\%\ ;\ N_{Tb^{3+}\ in\ solution} = 0.3L * 0.1\ M * N_A$$

Finally, plugging (6)-(8) into (5), we get the TA moles number:

$$(10)\quad n_{TA} \approx 3.42 * 10^{-4}\ mmol$$

In a racemic synthesis, the TA ee is initially:

$$(11) \qquad ee_{TA} = \frac{n_{D-TA} - n_{L-TA}}{n_{D-TA} + n_{L-TA}} = \frac{0.15*0.1 - 0.15*0.1}{0.3*0.1} = 0$$

When adding 1% seeds synthesized with pure D-TA, the TA ee will change to approximately:

$$(12) \qquad ee_{TA} = \frac{n_{D-TA} - n_{L-TA}}{n_{D-TA} + n_{L-TA}} \approx \frac{3.42*10^{-4}}{0.3*0.1 + 3.42*10^{-4}} = 0.01$$

This small change in ee cannot account for the large NC ee (~1), more so at lower seed %.

4. Discussion of the origin of the measured emission

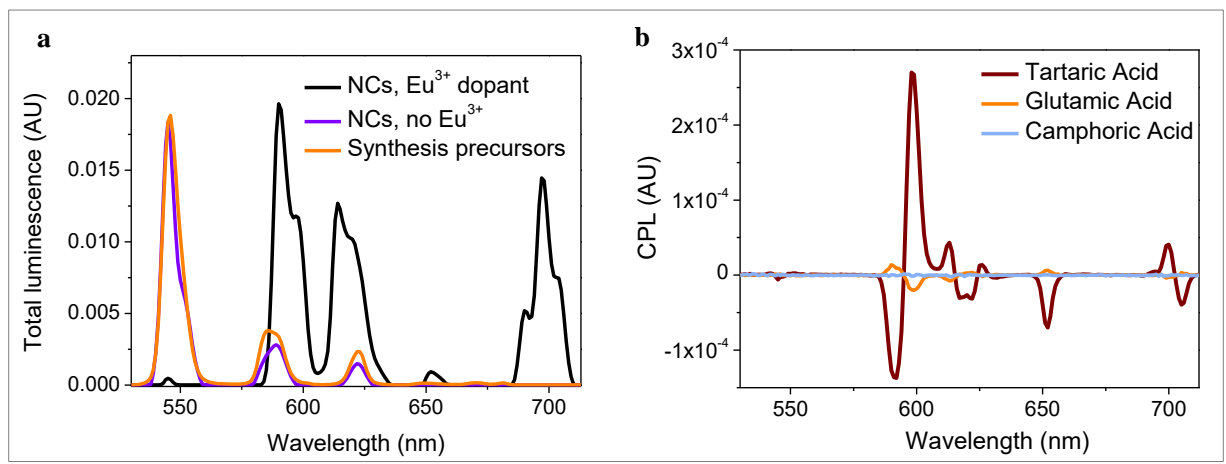

**Figure S3. a.** The luminescence spectra (excitation wavelength $\lambda_{ex} = 365$ nm) of : unreacted precursors for TbPO$_4$·H$_2$O synthesis with 5% Eu$^{3+}$ dopant, TbPO$_4$·H$_2$O NCs doped with 5% Eu$^{3+}$ and undoped TbPO$_4$·H$_2$O NCs. **b.** The CPL spectra (excitation wavelength $\lambda_{ex} = 365$ nm) of several NCs synthesized in the presence of different enantiopure ligands. All other reactions conditions were identical, and the final concentrations of the NCs were similar, verified by their luminescence.

It can be seen that in Eu$^{3+}$ doped TbPO$_4$·H$_2$O NCs the emission stems almost entirely from the Eu$^{3+}$ ions in the NCs. Doping with Eu$^{3+}$ adds several lines to the emission spectrum, which are assigned to the $^5D_0$->$^7F_j$ transitions[1] of the Eu$^{3+}$ ion. This is concurrent with a decrease in Tb$^{3+}$ emission, which is due to the energy transfer from Tb$^{3+}$ to Eu$^{3+}$ [2]. The unreacted precursors show the characteristic emission of Tb$^{3+}$ ions in water, and do not show any detectable Eu$^{3+}$ emission. This is due to the low Eu$^{3+}$ concentration, combined with its weak absorption at the excitation wavelength (365 nm) and quenching by water molecules[3,4]. Only when it is in the crystalline matrix, protected from solvent water and activated by energy-transfer from the Tb$^{3+}$, does the Eu$^{3+}$ emission increase to measureable levels.

The strongest indication that the surface does not contribute to the emission and that the origin of the CPL is from Ln$^{3+}$ ions inside the chiral NC is the fact that using different chiral molecules to direct the NC handedness results in the same CPL lineshape (Fig. S3b). If the surface had a considerable contribution, changing the chiral ligand, which is attached to the NC surface, would have changed the line shape of the CPL. The fact that the line shape is completely retained despite different ligands demonstrates that NCs of the same phase and crystal structure precipitate, and that the emission stems from the ions located at the NCs' interior. The difference between the spectra is in the magnitude of the signal, such that tartaric acid is the best handedness-directing ligand tested, and camphoric acid is the worst.

5. Notable $g_{lum}$ values of an enantiopure nanocrystal sample

For enantiomerically pure NCs synthesized with 100% D-TA:

| Wavelength [nm] | $g_{lum}$ |
|---|---|
| 590 | -0.05 |
| 598 | 0.16 |
| 612 | 0.06 |
| 621 | -0.03 |
| 651* | ≥0.4 |
| 699 | 0.03 |
| 704 | -0.04 |

**Table S1.** The dissymmetry factor ($g_{lum}$) for several CPL peaks

* The transition at 651 nm is electrically-forbidden, and thus its intensity is relatively low. This, combined with the large scattering of light from the NCs in solution, makes the determination of the exact $g_{lum}$ value at this wavelength problematic. We report the estimate for the lower boundary for $g_{lum}$ for this line.

6. <u>Enantiomeric excess of NCs with and without stirring</u>

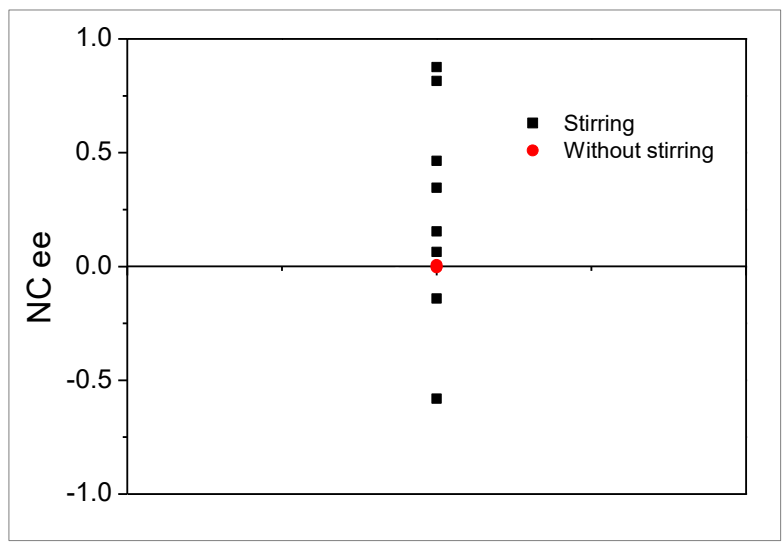

**Figure S4.** The enantiomeric excess of the NCs with and without stirring. All syntheses were carried out without TA (the handedness-directing molecule), and at 40°C, which is below the critical temperature for spontaneous symmetry breaking. The "no-stirring" synthesis (red circles) was repeated thrice.

The results clearly indicate that stirring is essential for the spontaneous symmetry breaking process. With neither stirring nor presence of TA, the ee of the NCs is either at the threshold of our CPL instrument noise level $(g_{lum} \sim 10^{-4})$ or below it.

7. <u>'Hysteresis'-like behavior at lower synthesis temperatures</u>

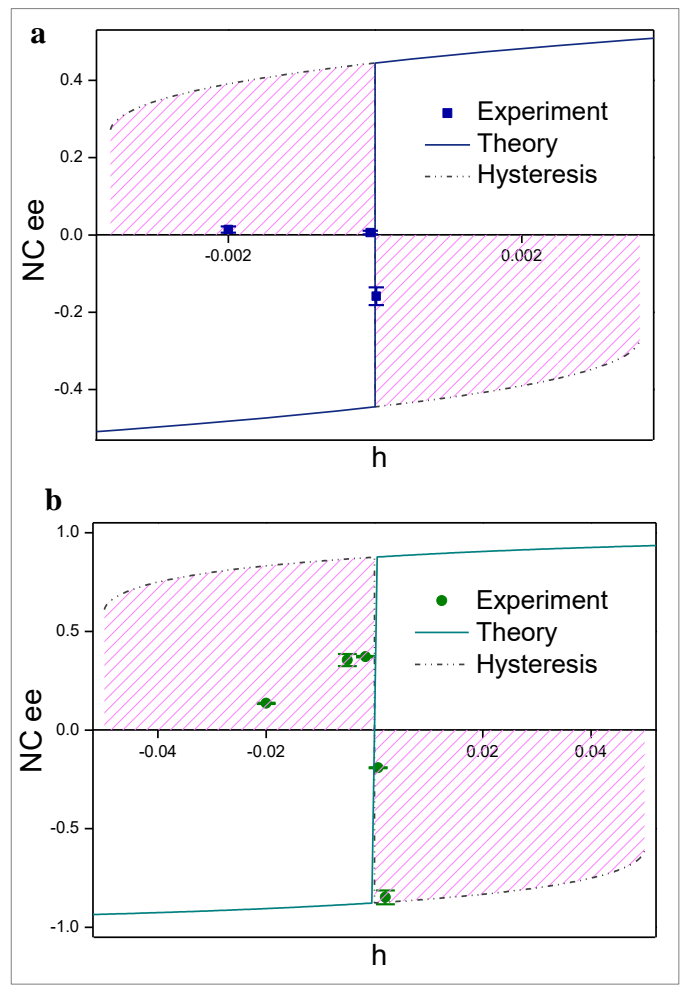

**Figure S5. The NC enantiomeric excess as a function of TA ee for syntheses with the 'opposite' expected handedness.** When the synthesis is performed with very low $h$ values, and below the critical temperature, then in some cases one may obtain NC ee that is opposite to the handedness dictated by the TA ee. Such cases would appear in the shaded quadrants in the plots above. Note the different scales in the two panels. This phenomenon is equivalent to "hysteresis" in other systems undergoing phase-transitions. The dashed lines at the borders of the shaded areas are calculated by the model used in this work.

# REFERENCES

1. Binnemans, K. Interpretation of europium(III) spectra. *Coord. Chem. Rev.* **295,** 1–45 (2015).

2. Di, W. *et al.* Photoluminescence, cytotoxicity and in vitro imaging of hexagonal terbium phosphate nanoparticles doped with europium. *Nanoscale* **3,** 1263–9 (2011).

3. Beeby, A. *et al.* Non-radiative deactivation of the excited states of europium, terbium and ytterbium complexes by proximate energy-matched OH, NH and CH oscillators: an improved luminescence method for establishing solution hydration states. *J. Chem. Soc. Perkin Trans. 2* **2,** 493–504 (1999).

4. Bunzli, J. G. & Eliseeva, S. V. *Basics of Lanthanide Photophysics. Springer Series on Fluorescence* (2011). doi:10.1007/4243